\documentclass[preprint,12pt,onecolumn]{aastex}

\begin{document}
\title{Determining Accurate Distances to Nearby Galaxies}
\author{Alceste Bonanos}
\affil{Harvard-Smithsonian Center for Astrophysics, 60 Garden St., 
Cambridge, MA 02138}

\begin{abstract}

We have undertaken several projects with the purpose of determining
accurate distances to nearby galaxies to calibrate the extragalactic
distance scale. Specifically, I describe the DIRECT project which aims
to derive the distance to M31 and M33 directly, using detached
eclipsing binaries and the Baade-Wesselink method for Cepheids. I also
present a ``hybrid'' method of discovering Cepheids with ground-based
telescopes using image subtraction and then following them up with the
{\em HST}\/ to derive Cepheid period-luminosity distances.

\end{abstract}

\section{Introduction}

Distances to extragalactic objects are known with an accuracy of less
than 10-15\%. This is due to the fact that standard candles available
to astronomers are not completely understood theoretically and most
importantly, that there are large uncertainties in the current anchor
galaxy of the extragalactic distance scale, the Large Magellanic Cloud
(LMC). Cepheids are examples of such distance indicators: the periods
of Cepheid variables are tightly correlated with their
luminosities. The correlation seems to depend on metallicity, but this
dependence is not well understood and is controversial. Also, the
distances to the LMC obtained with the same technique but different
calibrations disagree \citep[see][Figure 8]{Ben02}. The LMC has the
advantages of being nearby and easy to observe, however it introduces
problems as the anchor galaxy for the extragalactic distance
scale. The sources of systematic error associated with the LMC include
the differential reddening across the LMC, the elongation along the
line of sight, the metallicity of the galaxy and the zeropoint of the
Cepheid period-luminosity (PL) relation.

The uncertainty in the LMC distance not only translates into
uncertainty in the Hubble constant, but also in the calibration of
stellar luminosities and in constraining population synthesis models
for early galaxy formation and evolution. We therefore propose to use
other nearby galaxies, such as M31 and M33, as anchors of the
extragalactic distance scale bypassing the LMC and the systematic
errors associated with it.

\section{Hybrid Method for Measuring Distances}

We have proposed a ``hybrid'' approach for obtaining distances to
nearby galaxies with Cepheids \citep{Bon03}. Cepheids in nearby
galaxies can be discovered and characterized using large ground-based
telescopes and then followed-up with the {\em HST}\/ to obtain precise
distances. We demonstrated this by re-analyzing the excellent 8.2
meter VLT data of M83, obtained by \citet{Thi03}, using the image
subtraction method.

Blending must be taken into account in deriving the Cepheid distance
to nearby galaxies. For example, at the distance of M83 which is
$\sim4.5$ Mpc \citep{Thi03}, the median seeing of $0.76\arcsec$ of the
VLT data corresponds to $17\;$pc. As first discussed by Mochejska et
al. (2000), blending is the close association of a Cepheid with one or
more intrinsically luminous stars, which is the result of the higher
value of the star-star correlation function for massive stars, such as
Cepheids, compared to random field stars.  This effect cannot be
detected within the observed PSF by usual analysis. In M83, a large
fraction of the flux of a blended Cepheid could come from its
companions and would result in a significant distance bias.  The
discovery of Cepheids in nearby galaxies can be done adequately from
the ground given good signal-to-noise photometry; however, deriving
the Cepheid PL distance requires high spatial resolution {\em HST}\/
imaging.

With the image subtraction package ISIS \citep{Ala98,Ala00}, we were
able to detect 112 Cepheids, a nine-fold increase compared to the
number detected by \citet{Thi03} with the ``traditional'' method of
PSF photometry. We therefore demonstrate the power of image
subtraction, which should especially be used in crowded fields. These
additional Cepheids are valuable for determining the PL distance to
M83 accurately. However, {\em HST}\/ observations are necessary to
resolve blending effects.

\section{The DIRECT Project}

Starting in 1996 we undertook a long term project, DIRECT
(i.e. ``direct distances''), to obtain the distances to two important
galaxies in the cosmological distance ladder, M31 and M33. These
``direct'' distances will be obtained by determining the distance to
Cepheids using the Baade-Wesselink method and by measuring the
absolute distance to detached eclipsing binaries (DEBs). DEBs
\citep[for reviews see][]{And91,Pac97} offer a single step distance
determination to nearby galaxies and have the potential to establish
distances to M31 and M33 with an unprecedented accuracy of
5\%. However, DEBs are not easy to detect since they are intrinsically
rare objects (massive unevolved stars) and only certain configurations
produce eclipses. Now that large-format CCD detectors are available
and that CPUs are inexpensive, the DIRECT project has undertaken a
massive search for periodic variables, which is producing some good
DEB candidates.

We have so far analyzed observations taken with the 1.2 meter FLWO
telescope of six fields in M31, A-D, F \citep[][Papers I-VI]{Kal98,
Kal99, Sta98, Sta99, Moc99, Mac01} and recently field Y \citep[][Paper
IX]{Bon03b}. A total of 674 variables, mostly new, were found in M31:
89 eclipsing binaries, 332 Cepheids, and 253 other periodic, possible
long-period or non-periodic variables. We have analyzed two fields in
M33, A and B \citep[][Paper VI]{Mac01} and found 544 variables: 47
eclipsing binaries, 251 Cepheids and 246 other variables. Follow up
observations with the 2.1 meter KPNO telescope of fields M33A and M33B
produced 280 and 612 new variables, respectively \citep[][Papers VII,
VIII]{Moc01a,Moc01b}.

\begin{figure}[h]
\begin{center}
\includegraphics[width=0.6\textwidth]{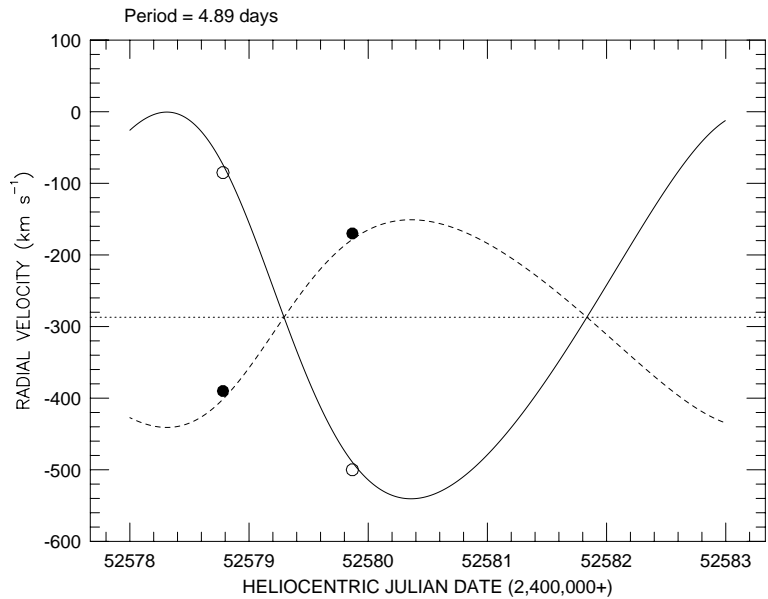}
\caption{Radial velocity curve for the 4.89 day period DEB M33A, from
two nights of data on Keck-II in October 2002.}
\label{fig}
\end{center}
\end{figure}

Of the $\sim 130$ eclipsing binaries, we have found 4 DEB systems
suitable for follow-up spectroscopy, 2 in M31 and 2 in M33. In October
2002, we obtained spectra of the two systems in M33 with ESI on
Keck-II but did not have enough phase coverage (see Figure~\ref{fig})
to derive the radial velocity amplitude accurately. However, we
concluded that M33A is a resolved double line eclipsing binary of
early B type that is suitable for distance determination and obtained
spectra with ESI on 3 more nights in September 2003. Deriving a radial
velocity curve is challenging, because early type stars have few
absorption lines in the visible spectrum, which are often broadened
and blended. We are currently analyzing these spectra and will soon
have the first direct measurement of the distance to M33.

We have also undertaken the first CCD variability study of the Draco
dwarf spheroidal galaxy with the FLWO 1.2 m telescope, producing 163
variable stars, 146 of which were RR Lyrae \citep{Bon04}. Using the
short distance scale statistical parallax calibration of \citet{Gou98}
for 94 RRab detected in our field, we obtained a distance modulus of
$\rm (m-M)_{0}=19.40 \pm 0.02\, (stat)\pm 0.15\,(syst)$ mag,
corresponding to a distance of 75.8 $\rm \pm 0.7 \,(stat) \pm 5.4
\,(syst)$ kpc to the Draco dwarf spheroidal galaxy.

\section{Summary}

The need for a new anchor galaxy or preferably for several anchor
galaxies to calibrate the extragalactic distance scale is long
overdue. The systematic effects introduced by using the LMC as the
anchor galaxy can be avoided now that 10-meter class telescopes have
become available. Large telescopes can be used for the detection of
Cepheids from the ground and later followed-up with the {\em HST}\/ to
obtain accurate distances, as demonstrated in M83. The DIRECT project
will determine geometric distances to M31 and M33 with an accuracy of
5\% with DEBs and the Baade-Wesselink method for Cepheids. Both of
these Local Group galaxies are excellent anchor galaxies for the
calibration of the extragalactic distance scale.

\end{document}